\documentclass[10pt,twoside,twocolumn,english]{article}
\usepackage[]{fontenc}
\usepackage[latin1]{inputenc}
\usepackage{amsmath}

\makeatletter

\providecommand{\LyX}{L\kern-.1667em\lower.25em\hbox{Y}\kern-.125emX\@}

\usepackage{epsfig}
\usepackage{ltexpprt}
\usepackage{latexsym}
\makeatletter
\let\ps@plain=\ps@empty
\makeatother

\usepackage{babel}
\makeatother
\begin{document}

\title{Determination of the Topology of a Directed Network\date{}}

\author{Darin Goldstein \cr Computer Science Department \cr Cal State Long
Beach \cr \textit{daring@cecs.csulb.edu}}

\maketitle
\begin{abstract}
We consider strongly-connected, \emph{directed} networks of identical
synchronous, finite-state processors with in- and out-degree uniformly
bounded by a network constant. Via a straightforward extension of
Ostrovsky and Wilkerson's Backwards Communication Algorithm\cite{OW},
we exhibit a protocol which solves the Global Topology Determination
Problem, the problem of having a root processor map the global topology
of a network of unknown size and topology, with running time $O(ND)$
where $N$ represents the number of processors and $D$ represents
the diameter of the network. A simple counting argument suffices to
show that the Global Topology Determination Problem has time-complexity
$\Omega (N\log N)$ which makes the protocol presented asymptotically
time-optimal for many large networks.
\end{abstract}

\section{Introduction}

\subsection{The Network Model. }

We consider strongly-connected \emph{directed} networks of identical
synchronous finite-state automata with in- and out-degree bounded
by a constant. These automata are meant to model very small, very
fast processors. The network itself has unknown topology and potentially
unbounded size $N$. Throughout this paper, we use the term {}``-port''
to refer to one of a number of unidirectional conduits through which
constant-size messages may pass from one processor to another. An
\emph{in}-\emph{port} to a processor will allow messages to flow unidirectionally
in towards the processor. \emph{Out}-\emph{port} is defined similarly.
We assume throughout that the number of in-ports and out-ports for
each processor is uniformly bounded above by a network constant $\delta \ge 2$.
The network is formed by connecting out-ports from processors to the
in-ports of other processors with wires. Not all in-ports or out-ports
of a given processor need necessarily be connected to other processors;
however, any given processor must have at least one connected in-port
and out-port. Note that even though the communication links are unidirectional,
a pair of processors is allowed to be connected with two communication
links, one in either direction, simulating a bidirectional link.

We assume that each processor in the network is initially in a special
``quiescent'' state, in which, at each time-step, the processor sends
a ``blank'' character through all of its out-ports. A processor remains
in the quiescent state until a non-blank character is received by
one of its in-ports.

The network has a global clock, the pulses between which each processor
performs its computations. Processors synchronously, within a single
global clock pulse, perform the following actions in order: read in
the inputs from each of their in-ports, process their individual state
changes, and prepare and broadcast their outputs.

The reason for modeling the processors by identical finite-state automata
is simple. In practice, many network protocols are expected to run
extremely fast. (One particular reason for this is that the network
topology or size might change if the protocol takes too long thereby
potentially rendering the computation obsolete. Obviously, if a processor
is randomly added or removed from the topology of the network in the
middle of the computation, a global topology determination is likely
to produce an incorrect result.) Commonly, a memory access can take
orders of magnitude longer than a simple state-change processor calculation.
It is therefore assumed that the processors involved will not have
time to access a large memory cache. The current technological trend
is to merge the memory functions that one generally associates with
a computing machine into the processor itself.

The protocol described below is presumed to begin when a certain processor
is signaled by some outside source. We call this processor the \emph{root},
and assume that every processor knows whether or not it is the root.
A protocol ends when the root enters a special terminal state indicating
that the computation has successfully completed. In our computational
model, we calculate the time-complexity of a protocol in terms of
the total number of global time steps between initiation and termination.
Of course, the aim is to minimize this time-complexity.

\begin{Definition}Throughout this paper, we will use $N$ to represent
the total number of processors in the network. $D$ will represent
the diameter of the directed network.\end{Definition}

\subsection{The Global Topology Determination Problem}

\subsubsection{Statement of the Problem. }

As previously mentioned, our computational model is designed to realistically
simulate a large network of small and fast processors with only the
capacity for reliable unidirectional communication. More specifically,
we have a large network of powerful computers each equipped with a
very fast communication processor (a communication device separate
from the main processor of the computer which is presumably engaged
in tasks other than simple communication protocols). These communication
processors are modeled by the finite-state automata. (We think of
finite-state machines as having a small/constant amount of memory
which allows them to work faster than a larger, more complex machine.)
One such computer initiates the protocol by nudging its communication
processor out of quiescence thereby forcing the device to take on
the role of the root. The protocol then commences via messages passing
between the various communication processors. At each step of the
protocol, the root is piping its computational transcript to the computer
to which it is attached. By the time the protocol has completed execution
(i.e. the root has re-entered quiescence after informing its master
computer that the algorithm has completed), the root's computer has
enough information in the form of input symbols read from its communication
device to reconstruct the global topology of the network.

One point should be clarified: When considering explicit protocols
on finite-state processors, an important characteristic of the processors
is the ability to recognize whether their in-ports and out-ports are
connected to other processors or not. The ability of a processor to
recognize whether its in-ports are connected to other processors is
called \emph{in-port awareness}. A similar definition applies to \emph{out-port
awareness}. When dealing with practical applications it is natural
to assume that processors have in-port awareness. (The quiescent (resting)
state of a processor is to constantly send out the blank character
$b$. If a processor is receiving $\emptyset $ (nothing) and not
$b$ through one of its in-ports, the processor knows immediately
that there cannot be a processor connected to that in-port.) However,
out-port awareness is not necessarily a given in some communications
systems. We will assume throughout this paper that processors have
both in-port and out-port awareness as did Ostrovsky and Wilkerson
in \cite{OW}.

\subsubsection{Previous Research\label{subsubsection:priorresearch}. }

Mapping the global network topology is an extremely important primitive
utilized for message routing as well as investigated for its intrinsic
theoretical interest. The literature on the subject of message routing
and topology mapping is immense. The most obvious example of practical
network topology determination is Internet mapping. Internet mapping
protocols are not in short supply. Mainwaring et al. in \cite{MCSW},
for example, present and prove the theoretical correctness of a protocol
for mapping the Myrinet system-area network at U.C. Berkeley. More
ambitiously, R. Govindan and H. Tangmunarunkit \cite{govindan00heuristics}
present and discuss a program called Mercator which performs {}``informed
random address probing'', a heuristic useful for mapping the entire
Internet topology. And perhaps most impressively, in \cite{burch99mapping},
Cheswick, Burch, and Branigan actually map the Internet topology almost
entirely. (In fact, they even perform an analysis of the NATO bombing
campaign on Yugoslavia's network connectivity.) Numerous other mapping
protocols and results can be found in these respective bibliographies.

Of course, the Internet is commonly considered a bidirectional network
with a fairly predictable overall topology (from a theoretician's
point of view). We will focus on \emph{directed} networks of totally
unknown topology and potentially unbounded size, and the processors
are assumed to be finite-state. The fact that our network assumptions
are so general make our results applicable to other, more specific
networks (e.g. any of the above references). However, our solution
might not be the most efficient from a practical point of view for
these other specific network types; this is especially true if one
has the foreknowledge that the network in question has particular
properties such as bidirectionality, greater processor memory, or
an easy-to-map hypercube topology. On the other hand, networks of
the more general kind occur more often than one might think (e.g.
GPS satellites, encrypted one-way radio military networks, bidirectional
networks with in-port or out-port shutdown failures at individual
processors, etc.). Network restrictions such as unidirectional communication
and finite-state processors make outlining a general topology mapping
protocol a nontrivial exercise. Luckily, Even, Litman, and Winkler's
{}``snake'' data structure \cite{ELW} (used in Ostrovsky and Wilkerson's
BCA in \cite{OW} and the equivalence proof in \cite{GM} and modified
for use in the protocol below) is virtually tailor-made for the purpose.

\section{Data Structures}

\subsection{Speed. }

The protocols about to be presented make use of several computational
constructs, each of which is assigned a certain characteristic ``speed.''
(The {}``speed'' concept can be referenced as far back as \cite{McMin}.)
This is \emph{not} to say that certain messages move faster through
the network than others. All computations and outputs are strictly
synchronous with respect to the global network clock.

In the protocol that follows, the speeds that we utilize are speed-1
and speed-3. The method by which we implement a speed is as follows:
A speed-1 construct will enter a processor. It will then remain there
for 3 global clock ticks. At the third clock tick, it will proceed
along its designated path. Similarly, a speed-3 construct will wait
only 1 global clock tick. Thus, in reality, the implementation specifies
that a speed-1 construct moves 3 times slower than a speed-3 construct.

\subsection{Tokens\label{subsection:tokens}. }

Tokens are the simplest data structure possible on networks of finite-state
machine. They should be thought of as markers that can be passed from
one processor to another via the edges of the network. The token concept
has been in use since the first solution to the Firing Squad Synchronization
Problem for the bidirectional line \cite{McMin}. The definitions
we give below of {}``breadth-first'' and {}``loop'' tokens were
first outlined in \cite{GM} though the behaviors were also utilized
in \cite{ELW,OW}. 

We employ two main varieties of tokens. \emph{Breadth-first tokens}
can be thought of as moving within a {}``breadth-first-search tree''
in the following sense: We arrange it so that each relevant processor
in the network has a {}``parent'' marker associated with one of
its in-ports. (The method by which various breadth-first-search trees
are constructed by snakes, as well as how each processor designates
its {}``parent'' in-port, is discussed in Section \ref{subsubsection:growingsnakes}.)
We then declare that a breadth-first token will only be accepted by
a given processor when either (a) the processor creates the token,
in which case the processor will not have a parent in-port, or (b)
the token comes through the processor's parent in-port. If a breadth-first
token is received through a non-parent in-port or by a quiescent processor,
it is ignored. Breadth-first tokens are passed out of every out-port;
thus, breadth-first tokens multiply in number as time goes on (as
long as they stay within the confines of the breadth-first-search
tree.) In summary, if a breadth-first token is created at the root
of its associated breadth-first tree, then $t$ time steps later there
will be a token at each processor that is a distance of $t$ from
the root, and none elsewhere. (If the tree has length less than $t$,
of course, there will be no tokens anywhere.) 

\emph{Loop tokens} travel along a specified marked loop within the
network. (How loops get marked is described in Section \ref{subsection:markedloops}.)
A processor on the loop that receives a loop token simply passes it
to the next processor on the loop. Thus, $t$ time steps after its
creation, a loop token will be $t$ processors away from the processor
that created it, along the marked loop. When any loop token reaches
it creator processor, it is absorbed (i.e., not sent around again). 

Note that tokens can only carry along with them a constant (very small)
amount of information since they are only of constant size. The next
data structure takes care of this problem.

\subsection{Snakes\label{subsection:snakes}. }

Our description of the snake data structure closely follows that in
\cite{GM}. The concept of a data-carrying \emph{snake} was invented
by Even, Litman, and Winkler in \cite{ELW}. Snakes are the solution
to the problem of the limited data-carrying capabilities of tokens.
A snake is capable of carrying an arbitrarily large amount of data,
but for this reason, it must reside in a collection of adjacent processors
rather than a single processor. 

A ``snake'' is a string -- which may be arbitrarily long -- made up
of an alphabet of $2(\delta ^{2}+\delta )+1$ characters, namely $\delta ^{2}+\delta $
head characters, $\delta ^{2}+\delta $ body characters, and a unique
tail character. (Recall that $\delta $ is a fixed constant of the
network.) The characters comprising the string are stored in adjacent
processors, one character per processor. These characters encode a
path by specifying a series of in- and out-ports. (Note that a token
could never do such a thing, since a path in the network can grow
arbitrarily long.)

We require two main snake types, which we call growing and dying.
\emph{Growing snakes} are used to generate encoded paths of the network,
and \emph{dying snakes} are used to mark encoded paths. Our protocol
requires two kinds of each of the two snake types; specifically, we
will need out-growing, in-growing, out-dying, and in-dying snakes.
{}``Out'' and {}``in'' are meant as a mnemonic; out-snakes are
generated at the root and proceed outward from it, while in-snakes
are generated elsewhere and trigger some action when they reach the
root. Out-growing, in-growing, out-dying, and in-dying snakes will
be referred to as OG-snakes, IG-snakes, OD-snakes, and ID-snakes in
what follows. 

Each of the four kinds of snake gets its own alphabet of characters
to describe it; this allows processors to determine with which kind
of snake they are dealing. We will spend a section on each type, elucidating
its respective behavior. First, we need to go over some general rules
common to all snake types.

\subsubsection{General Snake-handling Rules.}

\begin{itemize}
\item All snakes are speed-1. 
\item Snakes of different types do not interact. A processor can handle
different snake types simultaneously without getting confused because
snake types are distinguished by their alphabets. Note that this does
not impose arduous memory constraints upon the processors (which are
finite-state machines) since there is only a constant number of snake
types. 
\end{itemize}

\subsubsection{Growing Snakes\label{subsubsection:growingsnakes}.}

Growing snakes function as information generators. We define the \emph{initiator}
to be the processor from which the growing snakes first emanate. The
\emph{terminator} is defined to be the processor that the snakes are
attempting to reach. Growing snakes grow in a breadth-first manner;
the first growing snake to reach the terminator processor will have
encoded within its body a minimal-length path from the initiator to
the terminator. Upon reaching the terminator, a growing snake head
might then initiate some further action based on the protocol and
the state of the terminator. The rules for handling growing snakes
are outlined below; the rules for handling in-growing and out-growing
snakes are identical (just replace {}``IG'' with {}``OG''). We
assume that we are using in-growing snakes in most of the discussion
below for concreteness.

\begin{itemize}
\item First, the head characters of the baby growing snakes are generated
by the initiator. This processor sends an IG-snake head character
out of every out-port during the first time step. The particular head
character to be sent will correspond to the out-port from which it
is being sent. For every $i$ between 1 and $\delta $, the growing
head snake character $IGH(i,*)$ will be sent through out-port $i$.
When a processor receives any growing snake character with $*$ as
its second parameter (and this applies to body as well as head characters),
the processor notes the in-port $j$ through which the character arrived
and changes the $*$ to $j$. Thus, when the $IGH(i,*)$ is received,
it is changed to $IGH(i,j)$ where $j$ is the number of the receiving
in-port. During the next time step, the initiator will send a tail
character $IGT$ through every out-port. Thus a baby snake is born. 
\item When a processor receives an in-growing snake character (again, for
concreteness) for the first time, it marks itself $IG$-visited, and
marks the in-port through which the growing snake character was passed
as its $IG$-parent%
\footnote{If two or more IG-snakes arrive simultaneously, the one arriving through
the lowest-numbered in-port is deemed {}``first.''%
}. (These marks will be used later by certain breadth-first tokens;
see Section \ref{subsubsection:RCA}.) Only this first IG-snake will
be allowed to pass through the processor; all other IG-snake characters
will be ignored. Thus, an IG-snake will carve out a breadth-first-search
tree. Growing snake characters are periodically removed from the network.
Until this removal occurs, however, each growing snake carves out
a breadth-first-search tree. 
\item When a processor receives a non-tail IG-snake character, it simply
sends this character through all out-ports during the next time step.
Once the processor sends the character out, it need not retain it
in {}``memory.'' (In this way, the processor simply passes the head
and body through every out-port. Thus arbitrarily long paths can be
generated.) 
\item Once a processor receives the tail of an IG-snake, instead of simply
passing it through like the other body characters, the processor generates
a new body character. For all $i$ between 1 and $\delta $, it simultaneously
sends the character $IG(i,*)$ through out-port $i$; thus a new body
character is generated to mark the current processor's position in
the path. Only after this new character is passed along does the processor
send the tail through. Note that the $*$ is changed to reflect the
appropriate in-port when the body character is received by the next
processor in turn.
\end{itemize}

\subsubsection{Dying Snakes.\label{subsubsection:dyingsnakes}}

Dying snakes function as path markers. After a path is generated by
the growing snakes, it is the responsibility of the dying snake to
mark the generated path so that the processors on it will know (a)
that they lie along a special path and (b) which in-port and out-port
they should use for funneling information along the path. In our protocol,
ID-snakes will be formed by converting the characters of an OG-snake
into ID-snake characters as they pass through one particular processor;
OD-snakes will be created from ID-snakes in a manner to be described
in Section \ref{subsubsection:RCA}. The rules for handling ID-snakes
are outlined below; the rules for handling OD-snakes are identical
(just replace {}``ID'' with {}``OD''), except where noted. 

\begin{itemize}
\item An ID-snake will mark a path generated by an OG-snake (see Section
\ref{subsubsection:growingsnakes}); thus, since an OG-snake carves
out a breadth-first-search tree, the path will never self-intersect.
Similarly, neither will a path that an OD-snake is to mark. However,
the concatenation of the two paths (which, in our protocol, will always
be a loop that includes the root) may self-intersect; any processor
will appear at most twice on the concatenation. We will, eventually,
want to consider the concatenation as a whole; to this end, we imbue
each processor with two {}``predecessor in-ports'' (numbered 1 and
2) and two {}``successor out-ports'' (ditto). 
\item Whenever a processor receives the head character $IDH(i,j)$ of an
ID-snake, it sets predecessor in-port \#1 equal to the number of the
in-port through which it received the character, and sets successor
out-port \#1 equal to $i$. These two values indicate the two edges
of the path incident to the processor. OD-snakes work identically,
except that they use predecessor in-port \#2 and successor out-port
\#2. The head character is then discarded (not sent through any out-port). 
\item If the next ID-snake character that the processor receives through
the predecessor in-port is $ID(i',j')$, it gets sent through the
successor out-port as $IDH(i',j')$. (In other words, the next ID-snake
body character that comes through gets converted to the new head.)
The processor then passes all further ID-snake characters received
through its predecessor in-port to its successor out-port exactly
as received. If the next character happens to be a tail, then it gets
sent through the successor out-port as is. In our protocol, ID-snakes
will be converted into OD-snakes at the root; this will provide an
exception to these rules, for at the root all ID-snake characters
are converted into OD-snake characters instead. In addition, as might
be expected, the root will set predecessor in-port \#1 and successor
out-port \#2 appropriately as the dying snakes go through; its other
two ports will not be needed, as we will show in Section \ref{subsubsection:RCA}. 
\item An ID-snake only propagates along the path it is marking, and a maximum
of one will be in the network at any given time, so we need not worry
about ID-visited markings. 
\end{itemize}

\subsection{Marked loops.\label{subsection:markedloops}}

As mentioned in Section \ref{subsubsection:dyingsnakes}, we will
be using dying snakes to mark certain loops (not necessarily simple)
that include the root. We will refer to this structure repeatedly
throughout the paper, and thus make the following definition: 

\begin{Definition}

A \emph{marked loop} will be a loop marked by dying snakes in the manner
described in Section \ref{subsubsection:dyingsnakes}.  The root must be one
of the processors on the loop.  The loop may or may not be simple, but no
processor or edge can appear more than twice on it.

\end{Definition}

Each processor will have its predecessor and successor port (or, if
necessary, ports) set by the dying snakes. A processor with only predecessor
in-port \#1 set will only accept a loop token through that in-port;
it will subsequently pass the token through successor out-port \#1%
\footnote{Once again, the root will provide an exception to this rule; it will
accept a loop token only through predecessor in-port \#1, but will
pass it through successor out-port \#2;%
}. Similarly, a processor with only predecessor in-port \#2 set will
only accept a loop token through that in-port; it will subsequently
pass the token through successor out-port \#2. Finally, a processor
with both predecessor in-ports set will initially accept a given loop
token only through predecessor in-port \#1 (it will pass the token
through successor out-port \#1, of course); it then waits for the
token to come through predecessor in-port \#2 (at which point it passes
the token through successor out-port \#2); it then will expect the
next such loop token through predecessor in-port \#1 again. 

We will hereon refer to the predecessor in-port (resp. corresponding
successor out-port) through which a loop processor awaits a loop token
as the \emph{appropriate} predecessor in-port (resp. successor out-port).

\section{The Global Topology Determination Algorithm\label{section:algorithm}}

\subsection{Description of the Algorithm. }

In the discussion to follow, we will assume that two auxiliary protocols,
the Backwards Communication Algorithm (BCA) and the Root Communication
Algorithm (RCA), are available for use. The BCA is a method for sending
information {}``backwards'' along a unidirectional edge in the network,
and the RCA is a method for communication information from any given
node to the root. We defer more complete descriptions of each until
Sections \ref{sub:The-Backwards-Communication} (BCA) and \ref{sub:The-Root-Communication}
(RCA).

After initiation by its master computer, the root releases a DFS (Depth
First Search) token through its lowest-numbered connected out-port.
This token performs a depth first search of the entire network remembering
along the way through which out-port it has been most recently passed
and through which in-port it was most recently received. (The DFS
token is to be thought of as having the same basic structure as a
snake character with two entries where in-port and out-port labels
can be stored.) The information stored in this token is conveyed to
the root as the depth-first-search progresses.

We assume that the reader is somewhat familiar with the mechanics
of depth-first search on directed graphs. We will give a brief overview
of the depth-first search using finite-state processors. In the following
discussion, a ``forward edge'' refers to an existing edge of the network
representing a path along which messages are passed unidirectionally.
The reason we even bother making this distinction is that because
of the BCA, we have a method of passing information \emph{backwards}
through an edge. When a piece of information gets passed through a
legitimate edge of the network, we say that it gets passed \emph{forward}
through the edge.

To perform the depth-first search, any given processor, after receiving
the DFS token for the first time%
\footnote{If a processor is receiving the DFS token for the first time, it must
be through a forward edge of the network.%
}, notes the in-port through which it received the DFS token. The processor
also marks that in-port as its parent and then passes the DFS token
out its lowest-numbered connected out-port. After the processor gets
the DFS token back \emph{via the BCA%
\footnote{A processor may get the DFS token back through a forward edge of the
network after it already has marked its parent in-port. In that case,
the processor would use the BCA to send the DFS token back since a
processor never wants more than one parent.%
}}, it marks that out-port finished and sends the DFS token out of
the lowest-numbered unfinished connected out-port, and so on. When
all of the processor's out-ports are finally finished, the processor
sends the DFS token back through its parent in-port via the BCA. Once
the root has finished all of its out-ports, the depth-first search
is over.

The root is updated as to the progress of the protocol via the following.
Upon receipt of the DFS token, a processor initiates one of the following
two tasks. Once the task is completed, the DFS token is passed on
according to the rules of depth-first-search outlined above. (As indicated
above, whenever the DFS token needs to move backwards along an edge
of the network, it uses the BCA.)

\begin{itemize}
\item If the token was \emph{not} received through use of the BCA, the processor
performs the RCA using the FORWARD token. We assume that there are
$\delta ^{2}$ possible FORWARD tokens. The FORWARD token that gets
sent depends on which out-port sent the DFS token and which in-port
received the DFS token. For example, if the DFS token was passed out
of out-port 4 of one processor and into in-port 1 of another, then
FORWARD token $(4,1)$ is sent. (The FORWARD token has the same basic
structure as a snake character.)
\item If the DFS token was received via a backwards edge (i.e. if the token
was passed to the processor by the BCA), the processor performs the
RCA using the BACK token.
\end{itemize}
The algorithm terminates when the root has completed the depth first
search of the network (i.e. finished all of its out-ports).

What is the master computer's strategy for mapping the network given
the computational transcript of its communication processor at the
root? We will describe the strategy as if the computer were drawing
a topological map as the algorithm was proceeding. The computer always
keeps track of the processors in the network that have performed previous
RCA's, allocating them names as new processors are ``discovered''
by the algorithm. (Recall that the computer at the root has the ability
to assign processors unique names even though the communicating devices
at the processors themselves cannot.) It will keep a stack of processor
positions as well. When a processor performs an RCA with a FORWARD
token, the computer pushes it onto the stack. The processor at the
top of the stack then points to the processor that has most recently
performed a RCA. (If the root has just initiated the Global Topology
Determination Protocol then we consider the root itself as having
performed the last RCA; the stack will initially consist of only the
root.) Whenever an RCA is run, the computer notes the characters of
the IG-snake that passes through the root as it is converted to an
OG-snake (see Lemma \ref{lemma:pathcorrectness}). From the characters
of the IG-snake, the root computer can accurately map both the in-ports
and the out-ports of the canonical shortest path to the current processor
$A$, the processor running the RCA. Because the protocol is deterministic
and always produces the same canonical shortest path from any given
processor $A$ to the root and back again, the computer can tell whether
the current processor $A$ has already been marked on the map. If
it has not yet been marked on the map, the computer marks it and creates
a name for it. At the end of the RCA, the computer notes whether a
FORWARD or BACK token is being passed around the loop. If it is a
FORWARD token, then the computer should draw a directed arrow from
the top processor on the stack to the current processor $A$ through
the appropriate out-port and in-port. Afterwards, the computer pushes
processor $A$ onto the stack. (Recall that a FORWARD token indicates
that the depth first search has moved forward along an edge.) If it
is a BACK token, the computer simply pops the top processor off the
stack. Note that the top processor on the stack tracks the position
of the DFS token at any given point in time.

\section{Auxiliary Protocols and Correctness Proofs}

\subsection{The Backwards Communication Algorithm. \label{sub:The-Backwards-Communication}}

The Backwards Communication Algorithm (BCA) first appeared in \cite{OW}
and accomplishes the following: Assume there is a directed edge from
processor $A$ to processor $B$ in the network. The BCA is a method
for sending a message from processor $B$ to processor $A$ (\emph{backwards}
through the directed edge) such that $A$ gets the message, $B$ knows
when $A$ has gotten the message, and at the end of the transaction,
the rest of the graph is left undisturbed%
\footnote{It is important to note that the names $A$ and $B$ are just names
for the reader's convenience. Processors cannot all simultaneously
assign themselves unique names because of their finite-stateness.
Processor $B$ only recognizes Processor $A$ as the processor that
is on the other end of one of Processor $B$'s in-ports.%
}. The running time for each use of the BCA is $O(D)$.\\

\begin{Definition}

Given two distinct processors in the network, processor $A$ and processor
$B$, we define the \emph{canonical shortest path} from processor
$A$ to processor $B$ to be the unique path along which the first
growing snake released from processor $A$ that survives to reach
processor $B$ would travel.

\end{Definition}

\subsection{The Root Communication Algorithm. \label{sub:The-Root-Communication}}

In this section we will outline another algorithm, which we call the
Root Communication Algorithm (RCA), based on the idea of Ostrovsky
and Wilkerson's BCA, which we will use as part of the Global Topology
Determination Protocol presented in Section \ref{section:algorithm}.
The RCA accomplishes the following: processor $A$ communicates a
message to the root such that the root gets the message, processor
$A$ is aware of the completion of the algorithm, the computer at
the root is able to reconstruct the sequence of in-ports and out-ports
along the canonical shortest paths leading from the the root to processor
$A$ and from processor $A$ to the root, and at the end of the transaction,
the graph is left undisturbed. This auxiliary algorithm will be used
to send one of the two signals FORWARD or BACK to the root and to
allow the root computer to track the movement of the DFS token. The
Global Topology Determination Protocol is guaranteed to only be running
the RCA at a single processor at any given time. Throughout this section,
we will assume that processor $A$ is the processor that wishes to
communicate with the root.

\subsubsection{The Steps of the Root Communication Algorithm\label{subsubsection:RCA}. }

For the sake of brevity, we will abbreviate the snake types.

\begin{enumerate}
\item \label{item:IG}Processor $A$ becomes aware that it wishes to communicate
with the root (via a process outlined in the steps of the Global Topology
Determination Algorithm in Section \ref{section:algorithm}) and sends
IG-snakes to search for the root. Any processor receiving an IG-snake
character for the first time marks itself as ``IG-visited'', thus
preventing any subsequent IG-snakes from entering it. It will also
designate the in-port from which it received the IG-snake as its {}``IG-parent''
in-port. These markings will not be cleared until the release of KILL
tokens (in step \ref{item:KILL}); hence, the IG-snakes carve out
an IG-breadth-first-search tree.
\item \label{item:OG}Upon receipt of the head of the first IG-snake to
reach it, the root performs two actions simultaneously. First, the
root closes itself off to all other IG-snakes, ignoring any that attempt
to enter. The root will accept no further IG-snakes during this execution
of the algorithm. Second, the root begins to convert the IG-snake
to an OG-snake which it broadcasts out all out-ports. (To \emph{convert}
an IG-snake to an OG-snake, the root simply converts the IG-snake
characters it receives as input to OG-snake characters when they are
sent out.) When the root receives the tail of the IG-snake that it
is converting, it simply holds onto the tail and continues growing
the OG-snake normally: The root holds onto the tail character while
it sends out the character $OG(i,*)$ out of each of its out-ports
for every $i$ between 1 and $\delta $. Only after this character
is broadcast does the root send out the tail of the snake as $OGT$.
OG-snakes leave {}``OG-visited'' and {}``OG-parent'' markings
(and thus create an OG-breadth-first-search tree) similar to the IG-snakes
discussed in Step \ref{item:IG}. However, the OG-snakes do not respect
the IG-breadth-first-search tree and are therefore guaranteed to make
it back to processor $A$.
\item \label{item:marking}When processor $A$ receives the first OG-snake
head that survives, processor $A$ closes itself off to any subsequent
OG-snakes and converts the OG-snake to an ID-snake (note that processor
$A$ must therefore eat the head character of the OG-snake as if it
were an ID-snake character, then send the rest of the snake through
the appropriate out-port). The ID-snake then marks the path from processor
$A$ to the root. Eventually the root receives the head of an ID-snake
and converts it to an OD-snake as previously described. This OD-snake
then marks the path from the root back to processor $A$ which will
only receive the tail character $ODT$. At this point in the protocol,
there is a marked path in the network from processor $A$ to the root
and back again.
\item \label{item:KILL}As soon as processor $A$ receives the tail of the
OD-snake, processor $A$ performs two tasks simultaneously. First,
it releases a speed-3 breadth-first KILL token. The function of the
KILL token is to completely eradicate all traces of growing snake
characters in the network; both IG- and OG-snake characters and markings
are erased upon contact with a KILL token. KILL tokens are ignored
by those processors that do not have any growing snake markings or
characters in their memory. (KILL tokens do not affect the marked
path.) Second, processor $A$ releases a speed-1 loop token. This
token will either be a FORWARD or BACK token depending on the current
state of the network. Upon reception of the FORWARD/BACK token, processor
$A$ is guaranteed that one time step later, there will be no further
growing snake characters or KILL tokens percolating uselessly through
the network.
\item \label{item:terminate}Processor $A$ finally releases a speed-3 UNMARK
token around the marked path. Each processor the old marked loop,
upon receiving the token through its appropriate predecessor in-port,
passes the token through the appropriate successor out-port, then
forgets those predecessor and successor designations. Upon reception
of this UNMARK token, the root reopens itself to IG-snakes. After
the token makes it all the way around the marked path back to processor
$A$, processor $A$ reopens itself to OG-snakes and terminates the
algorithm.
\end{enumerate}

\subsubsection{Proof of Correctness of the RCA. }

In this section, we prove the non-trivial claims made about the RCA
presented in Section \ref{subsubsection:RCA}.

\begin{lemma}\label{lemma:pathcorrectness}

The root's master computer is able to determine the canonical shortest
paths leading to and from processor $A$ by the completion of the
algorithm.

\end{lemma}

\begin{proof}

Note that the canonical shortest path from processor $A$ to the root
is unique and is encoded in the body of the first (and only) IG-snake
to safely reach the root. Thus, to track this path, the master computer
can simply read off the in-ports and out-ports encoded in the body
of the IG-snake as it is converted to an OG-snake in Step \ref{item:OG}.
Similarly the canonical shortest path from the root to processor $A$
is also unique and is encoded in the body of the ID-snake that reaches
the root in Step \ref{item:marking}. The master computer can again
simply read off the relevant in-ports and out-ports as the ID-snake
is converted to an OD-snake. \end{proof}

\begin{lemma}\label{lemma:eradication}

After processor $A$ terminates the algorithm in Step \ref{item:terminate},
the network is left completely undisturbed by any data construct created
by the algorithm (snake characters/markings, tokens, etc.).

\end{lemma}

\begin{proof}

Because KILL tokens travel three times faster than snakes, it is obvious
that the tokens will eventually catch up with and eliminate the growing
snakes. To see that the KILL tokens will catch up precisely when claimed
(see step \ref{item:KILL}), note that if $L$ is the length of the
current marked loop in the network, then the snake heads have at most
a $2L$ head start. This implies that the KILL tokens will catch up
with the snake heads after a speed-1 token (i.e. the FORWARD/BACK
token) makes it around the loop once.\end{proof}

\begin{lemma}

Each execution of the RCA by any given processor $A$ takes time $O(D)$.

\end{lemma}

\begin{proof}

By inspection of the steps, the running-time is proportional to the
length of loop marked by the algorithm: $d(A,root)+d(root,A)$. This
quantity is itself trivially $O(D)$. \end{proof}

\subsection{Correctness Proof of the Global Topology Determination Algorithm}

\begin{lemma}

The Global Topology Determination Algorithm terminates in time $O(ND)$.

\end{lemma}

\begin{proof}

Each processor in the network performs at most $\delta $ RCA's and
at most $\delta $ BCA's. The running time for each of these subalgorithms
is $O(D)$ and the result follows. \end{proof}

\begin{theorem}

The computer at the root of a network performing the Global Topology
Determination Algorithm accurately maps the given directed network.

\end{theorem}

\begin{proof}

First, we claim that any time a FORWARD token is noted by the root
computer, the two processors between which it draws the directed arrow
have both already been mapped. Every time an RCA is run, processor
$A$ must be mapped before the token is sent out because a IG-snake
must be sent through the root before a FORWARD token. The previous
processor (i.e. the processor on top of the stack) has already been
mapped by a previous execution of the RCA and thus a FORWARD token
can always be traced between two processors already on the network
map at the time it is sent. Now we note that the DFS token must be
sent forward through every edge of the network and hence a FORWARD
token token is sent for every edge of the network. Thus all edges
get accurately mapped. \end{proof}

\section{The Lower Time Bound}

\begin{lemma}\label{lemma:graphs}

Let $G(N)$ be the number of bounded-degree strongly connected networks
of $N$ processors and diameter less than or equal to $2\log N+1$
with distinct topologies. Then there exists some constant $C$ such
that, for large enough $N$, $G(N)\geq N^{CN}$. (i.e. There are a
great many networks with small diameter.)

\end{lemma}

\begin{proof}

We only present a quick justification and leave the interested reader
to fill in the details. Consider the family of networks that consist
of a full binary tree emanating from a single node with bidirectional
edges (i.e. unidirectional edges in both directions) with a simple
loop that includes every processor on the bottom level of the tree.
Note that all such networks are of bounded-degree and strongly connected.
Every rearrangement of the processors included in the loop on the
bottom levels yields a distinct topology. A simple counting argument
suffices to complete the proof. 

\end{proof}

We make the convention that the processors' input/output set is called
$I$. The number of elements of the set $I$ is $|I|$.

\begin{lemma}\label{lemma:transcript}

For any given algorithm, after $x$ global clock ticks, the root can
have had one of a maximum of $|I|^{\delta x}$ possible computational
transcripts.

\end{lemma}

\begin{theorem}\label{theorem:lower}

Any algorithm which solves the Global Topology Determination Problem
has a time-complexity lower bound of $\Omega (N\log N)$.

\end{theorem}

\begin{proof}

Fix an algorithm which solves the Global Topology Determination Problem.
Let us assume that the algorithm terminates on graphs with $N$ processors
in less than or equal to $T(N)$ global clock ticks.

In order for the root to distinguish between different global topologies,
for any given network size $N$, there must be at least as many computational
transcripts as there are distinct network topologies. Otherwise, by
the pigeonhole principle, two distinct network topologies would have
to be distinguished by exactly the same computational transcript which
is, of course, impossible.

By Lemma \ref{lemma:graphs}, we know that for large enough $N$,
there exists a constant $C$ such that the number of distinct network
topologies is greater than or equal to $N^{CN}$. By Lemma \ref{lemma:transcript},
the number of possible computational transcripts the root can have
had is at most $|I|^{\delta T(N)}$. Thus we get: 

\[
|I|^{\delta T(N)}\geq N^{CN}\Rightarrow T(N)=\Omega (N\log N)\]
 \end{proof}

\bibliographystyle{plain}
\bibliography{optimal}

\end{document}